\documentclass[aps,amsfonts,prb,twocolumn,showpacs]{revtex4-1}

\usepackage{subfigure}
\usepackage{textcomp}
\usepackage{graphicx}
\usepackage{amssymb}
\usepackage{amsmath}
\usepackage{bm}
\usepackage{amsmath, amsthm, amssymb}
\usepackage{dsfont}
\usepackage[colorlinks=true,linkcolor=blue,urlcolor=blue,citecolor=blue]{hyperref}
\usepackage{url}

\def\beq{\begin{equation}}
\def\eeq{\end{equation}}
\def\bea{\begin{eqnarray}}
\def\eea{\end{eqnarray}}

\begin{document}
%
%

\title{Quantum percolation of monopole paths and the response of quantum spin ice}

\author{Matthew Stern$^{1,2}$, Claudio Castelnovo$^{3}$, Roderich Moessner$^{4}$, Vadim Oganesyan$^{2,5}$ and Sarang Gopalakrishnan$^{2,5,6}$}
\affiliation{$^1$ Department of Physics and Astronomy, Stony Brook University, Stony Brook NY 11790, USA \\
$^2$ Initiative for the Theoretical Sciences, Graduate Center, CUNY, New York, NY 10016, USA\\
$^3$ TCM Group, Cavendish Laboratory, University of Cambridge, Cambridge CB3 0HE, UK \\
$^4$ Max-Planck-Institut f\"{u}r Physik komplexer Systeme, 01187 Dresden, Germany \\ 
$^5$ Department of Engineering Science and Physics, College of Staten Island, Staten Island, NY 10314, USA \\
$^6$ Department of Physics, The Pennsylvania State University, University Park, Pennsylvania 16802, USA}
%
%

\begin{abstract}

We consider quantum spin ice in a temperature regime in which its response 
is dominated by the coherent motion of a dilute gas of monopoles through an 
incoherent spin background, taken to be quasi-static on the relevant 
timescales. 
The latter introduces well-known blocked directions that we find sufficient 
to reduce the coherent propagation of monopoles to quantum diffusion. 
This result is robust against disorder, as a direct consequence of the ground-state degeneracy, which disrupts the quantum interference processes needed for weak localization. 
Moreover, recent work [Tomasello \emph{et al.}, Phys.~Rev.~Lett.~\textbf{123}, 067204 (2019)] has shown that the monopole hopping amplitudes are roughly bimodal: for $\approx 1/3$ of the 
flippable spins 
surrounding a monopole, these amplitudes are extremely small.
We exploit this structure to construct a theory of quantum monopole motion in spin ice. 
In the limit where the slow hopping terms are set to zero, the monopole wavefunctions appear to be fractal; we explain this observation via mapping to quantum percolation on trees. 
The fractal, non-ergodic nature of monopole wavefunctions manifests itself in the low-frequency behavior of monopole spectral functions, and is consistent with experimental observations.

\end{abstract}

\maketitle
%
%

\section{\label{sec:intro}
Introduction
             }
Topological quantum spin liquids feature fractionalised quasiparticles~\cite{Wilczek2009,Balents:2010,Knolle2019,Broholm2019}, whose properties stem from the topological nature of the ground state~\cite{Fennell2009,KAdowaki2009,Morris2009,Han2012,Shen2016,Paddison2017,Hermanns2018}. 
It is tempting to regard these quasiparticles as simply dressed free particles~\cite{pwabook}; however, the fact that they are defined above a nontrivial and sometimes degenerate vacuum can qualitatively invalidate this free-particle picture. A case in point is quantum spin ice (QSI)~\cite{Gingras2014}. Classically, the ground state of spin ice has extensive entropy, as all configurations satisfying the ``ice rules'' are ground states~\cite{Harris:1997}; quantum fluctuations lift this degeneracy and give rise to quantum spin-liquid behavior~\cite{Moessner2003,Hermele2004,Balents:2010}.
However, while classical spin ice is well understood~\cite{Castelnovo2012}, we lack a definitive quantum counterpart (see e.g., Refs.~\onlinecite{Sibille2018,Tokiwa2018}). Potential probes of quantum spin ice often focus on its quasiparticles, which are expected to resemble quantum electrodynamics~\cite{Moessner2003,Hermele2004}.
Understanding
excitations and their interplay in a strongly correlated three dimensional
quantum spin system is, however, technically challenging (see e.g., Refs.~\onlinecite{Chen2017a,Chen2017b,Huang2018,Udagawa2019,Szabo2019}).

\begin{figure}[tb]
\begin{center}
\includegraphics[width = 0.48\textwidth]{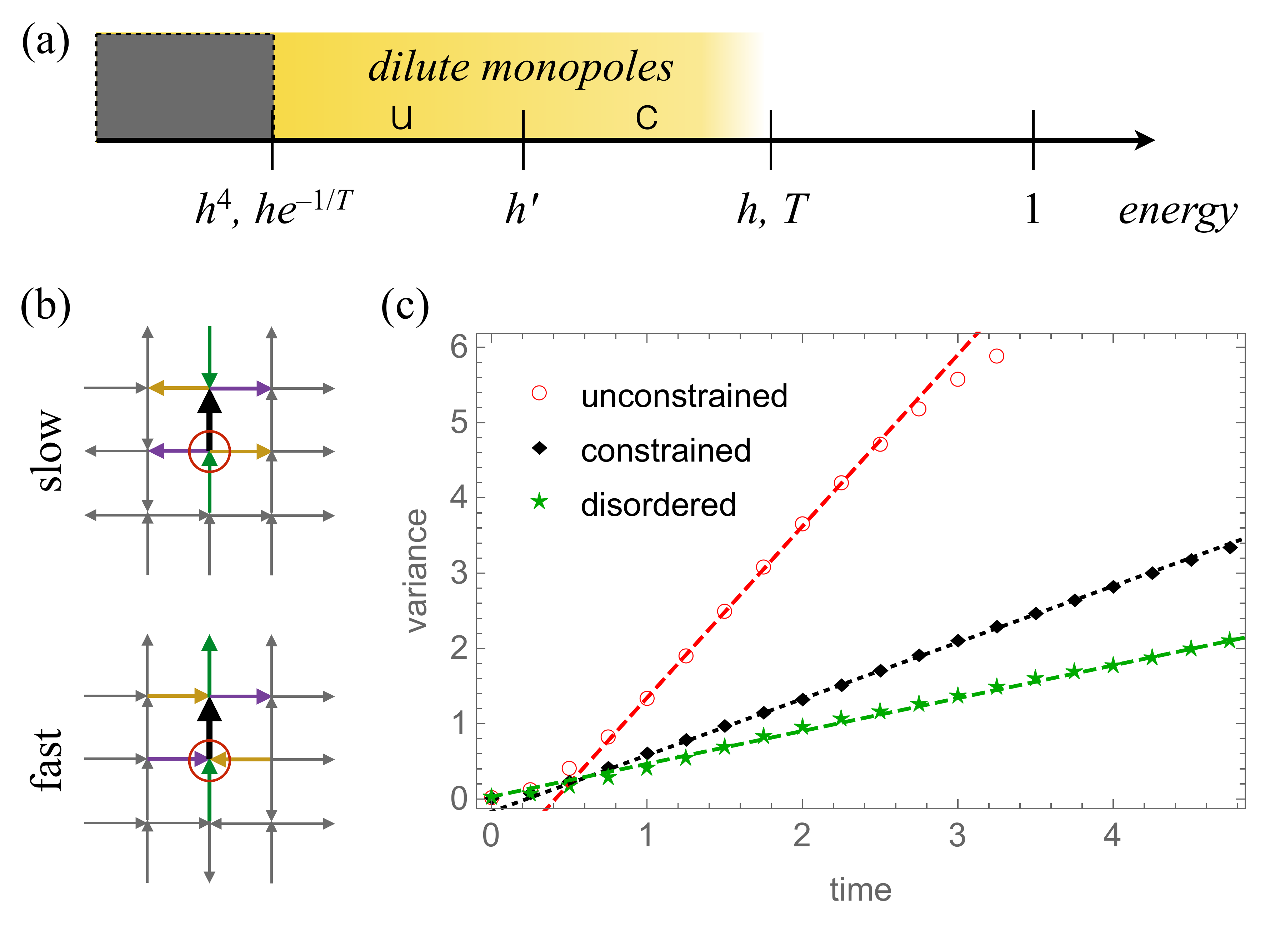}
\caption{(a)~Hierarchy of energy (or equivalently time/temperature) scales in QSI. The spin ice constraint acts on energy scales below the interaction strength (set to unity); the density of monopoles is exponentially suppressed in $T \ll 1$ and they hop on scales set by the transverse field $h$. On timescales $1/h \ll t \ll 1/h'$ (where $h'$ is the ``slow'' hopping), monopoles effectively experience \emph{constrained} motion; on longer timescales, this constraint is absent. At very long times, dynamics in the ground-state manifold, and inter-monopole interactions, are important. (b)~Constrained square ice model; open circles mark monopole positions. Consider the bold spins; fast (slow) configurations are those in which this spin can (cannot) move under $h$. The spins surrounding it are grouped into three inequivalent color-coded types; for the bold spin to be static each inequivalent pair must point oppositely~\cite{Tomasello2019}. (c)~Monopole diffusion in three cases: unconstrained motion, constrained motion ($h' = 0$), and constrained + disordered hopping, with $h \in [\frac{2}{3}, 1]$. The system sizes are $10^2$ for the unconstrained case, and $14^2$ for the constrained cases; there are no finite size effects out to the times shown. The diffusion constant is $1.15 \pm 0.03$ (unconstrained), $0.38 \pm 0.01$ (constrained), and $0.22 \pm 0.01$ (disordered).}
\label{fig:mainfig}
\end{center}
\end{figure}

This motivates the development of effective models of quasiparticle dynamics. In spin ice, the quasiparticles are not free pointlike excitations, but they are inextricable from the rearrangement of the underlying quantum spin state. The situation is particularly transparent at low but nonzero temperatures, when the underlying spin state is a \emph{statistical mixture} over all classically allowed ground states~\footnote{The quantum case when a superposition of classically allowed tensor product states forms is also very interesting, as recently explored for example in Ref.~\onlinecite{Morampudi2019}.}. When a monopole moves through this background, it leaves a trail of flipped spins behind; the presence of this observable trail prevents interference between different paths and leads to quantum diffusive behaviour as in the motion of holes in the Hubbard model~\cite{Brinkman1970,Mohan1991,Starykh1996,Chernyshev1999,Carlstrom2016}. Moreover, as we will see below, the monopole traverses a background with randomness that is quenched \emph{on the monopole hopping timescales}. This quenched randomness affects monopole motion by energetically blocking some directions and, as recently pointed out in Ref.~\onlinecite{Tomasello2019}, by suppressing the amplitude for certain monopole moves. 
We note in passing that an incoherent spin background can be induced not only by finite temperature but also by the proliferation of vison excitations (i.e., ring echange spin flip processes) at low but non-zero temperatures, for example close to a phase transition out of the QSI phase~\cite{Wang2021}. 

Here, we construct a theory of monopole dynamics in this intermediate temperature regime. The quenched randomness in hopping amplitudes, arising from the random background~\cite{Tomasello2019}, allows us to draw a connection between the dynamics of spin ice and quantum percolation, and, remarkably, enables more efficient numerical simulations (by only including the parts of Hilbert space that the monopole can visit). The resulting model is interesting in its own right: it presents a concrete setting where one can study kinetic constraints in quantum dynamics, and thus has affinities with certain models of disorder-free many-body localization~\cite{schiulaz2014, drh2014, garrahan_TI_MBL, Smith2017, Smith2018}. 
We find that the blocked directions in the underlying state reduce the coherent propagation of monopoles to quantum diffusion. 
Diffusion is robust against disorder, as a direct consequence of the ground-state degeneracy, which disrupts the quantum interference processes needed for weak localization.
Thus the bimodal distribution of transverse kinetic terms, or even explicit randomness in the hopping matrix elements, affects dynamics only quantitatively by decreasing the diffusion constant. 

With the more efficient numerical simulations, we are able to extract statistical properties of the monopole wavefunctions and their spectra. Remarkably, although transport is diffusive, the wavefunctions are not ergodic but multifractal. 
A fraction of states, meanwhile, are confined to move on finite spin clusters, resulting in characteristic sharp finite-frequency peaks in the density of states, conductivity and other spectral functions, in the absence of disorder. 
We account for these results by mapping monopole dynamics to a percolation problem on the Bethe lattice. 
We also briefly discuss the optical conductivity in this single-monopole limit, which has qualitatively similar features to those noted in Ref.~\onlinecite{pan2016} in experiments on Yb$_2$Ti$_2$O$_7$. 

Our results are relevant to ultrafast spectroscopy measurements
on QSI materials~\cite{pan2016}, and also to QSI implementations on quantum annealers~\cite{King2021}; 
moreover, the observation that some monopoles are confined to move on finite spin clusters 
potentially has consequences for inelastic properties and magnetic noise, which are accessible using neutron scattering~\cite{Broholm2019} and SQUID~\cite{Dusad2019,Watson2019,Tennant2021} measurements, respectively. 
%
%

\section{\label{sec:models}
Models and methods
             }
The phenomena relevant here are captured by the following effective nearest-neighbor Hamiltonian for spins $s > 1/2$ on the pyrochlore lattice: 
\beq\label{hmicro}
H = -J \sum\nolimits_{\langle ij \rangle} \mathbf{S}_i \cdot \mathbf{S}_j  - \Lambda \sum\nolimits_i (\mathbf{n}_i \mathbf{\cdot S}_i)^2 
\, , 
\eeq
which encompasses a nearest neighbour spin-spin interaction and a strong local easy-axis anisotropy~\footnote{Similar conclusions hold for more realistic microscopic QSI Hamiltonians, as discussed in Ref.~\onlinecite{Tomasello2019}}. The spins live on the sites of the pyrochlore lattice, and the local easy axes $\mathbf{n}_i$ point from the centres of one sublattice of tetrahedra to the centres of the other. There are four inequivalent directions. 

For $\Lambda \gg J$, each spin can be projected onto a ground-state doublet along its local easy axis; interaction terms parallel to this axis couple directly to the doublet, whereas transverse terms induce matrix elements involving the excited state and are therefore virtual processes, suppressed by factors of $1/\Lambda$. 
Since the easy axes of neighbouring spins are not parallel to one another, the exchange field 
\beq\label{hint}
\mathbf{h}_i = J \sum\nolimits_{j : \langle ij \rangle} \mathbf{S}_j 
\, , 
\eeq
has in general both a component parallel to the local axis $\mathbf{n}_i$, as well as a component tranverse to it. 
The longitudinal component fixes the ground state manifold. 
The transverse component---comprising terms $\sim S^z_i S^x_j$---gives dynamics to the spins~\footnote{Exchange interactions in generic Hamiltonians for spin systems on the pyrochlore lattice are more complex than Eq.~\eqref{hmicro}~\cite{Onoda2011}. These details do not affect our discussion; note that the results in Ref.~\onlinecite{Tomasello2019}, which we build upon here, were in fact obtained with the full-fledged exchange Hamiltonian.}. 
Both components depend on the (classical) configuration of the surrounding spins. In the ground state, a dominant longitudinal term pins the spin to its lowest energy state. 
In our work we are concerned with the dynamics of a single monopole in a regime where creation and annihilation events are energetically forbidden~\footnote{Monopoles must strictly be created in pairs; we artificially pin the other (anti-)monopole at a given location on the lattice, as separated as possible from the monopole whose motion we study.}. 
When a tetrahedron hosts a monopole (e.g., a three-in one-out defect in the ground state), three of the surrounding spins are free to flip, in which case the monopole hops to an adjacent tetrahedron at no energy cost: the longitudinal term correspondingly vanishes. The fourth spin cannot be flipped without introducing further violations of the ice rules, and its motion is energetically blocked. 
The three flippable spins are able to precess under the action of the transverse component. As shown in Ref.~\onlinecite{Tomasello2019}, the transverse component can take either a finite value (on average, in $2/3$ of the cases), or it can vanish ($1/3$ of the cases). 
In the latter case, even if there is no energy barrier preventing the spin to flip, there is also no matrix element inducing any dynamics and the spin remains static. 

We make the further simplification of working on a two-dimensional version of spin ice called square ice (equivalently known as the six vertex model). The criterion for a spin to be static is chosen in resemblance to the 3D case (as explained in App.~\ref{app:trees}), so that the average number of dynamical (neither blocked nor static) spins around the monopole is two. We expect the phenomena studied in this paper not to differ significantly between the two lattices, as they are \emph{locally} both trees with identical connectivity and---as we discuss below---loops do not play an important role in the dynamics~\cite{Udagawa2019}. 

We construct and diagonalize this model as follows. First we generate a random ice-rule obeying configuration with open boundary conditions; then we flip a single spin to create a pair of monopoles, and we move one member of this pair to the edge of the sample. Next we construct a tree graph of all configurations that can be reached from this initial configuration by moving the other monopole across all allowed spins. We keep track of the real-space position of the monopole at each node on this tree; however, inequivalent paths that reach the same node are treated as distinct, since they are physically distinguishable via the trail of flipped spins the monopole leaves behind. This remark is essentially exact out to a long timescale (compared to the characteristic time scale to flip a spin), on which nontrivial interfering paths---involving flipping the same trail of spins---can be constructed (see App.~\ref{app:trees}). These paths do not appear for the times and sizes we consider, and are quite fine-tuned; therefore we expect that any weak localization corrections they generate will be 
small. 

We study two limits: one in which only the blocked spins are prevented from flipping and all others have the same flipping amplitude; and another in which all static spins are treated as completely frozen and prevented from flipping. We call these two cases unconstrained and constrained, respectively. 
In the first case, the tree is regular and has connectivity $3$. 
When the static spins are frozen, each node of the tree (other than the first) can either have three allowed directions (branching), two allowed directions (linear), or one allowed direction (dead end). 
Eventually we stop this process, either because we have reached all accessible nodes, or because the Hilbert space of the tree becomes intractably large. 
The constrained case is not only more realistic, but also offers a numerical advantage, since the Hilbert space grows more slowly with the distance traveled, and therefore reliable simulations are possible to longer times. 
When non-vanishing, the magnitudes of the transverse components of the exchange fields from Eq.~\eqref{hmicro} are uniform across the system. 
We also consider the effect of adding static disorder to the transverse fields. 
%
%

\section{\label{sec:diffusion}
Diffusion
             }
As a first diagnostic tool, we look at the variance in the displacement of a monopole, $\langle x^2 \rangle$, vs. time $t$ (see Fig.~\ref{fig:mainfig}). 
We note that the direction of monopole motion is opposite to the direction in which the net magnetization of the system changes. Thus the monopole diffusion constant is directly related to the autocorrelation function of the total magnetization, as measured, e.g., spectroscopically~\cite{pan2016}. 

The blocked directions alone lead to purely self-retracing monopole paths, and therefore to quantum diffusion. This is seen as a linear growth of the variance $\langle x^2 \rangle = 2 D t$. Further constraints (e.g., ignoring the static spins) change the results only quantitatively (by reducing the diffusion constant $D$), similarly to the effects of disorder. 
The diffusion constant in the constrained case is roughly a third of what it is in the unconstrained case; disorder suppresses it yet further. 
This diffusion constant is related to the low-frequency limit of the monopole ``conductivity'' by the Einstein relation. 
%
%

\section{\label{sec:percolation}
Mapping to quantum percolation on trees
             }
Although the variance of the position grows linearly in time, in the constrained case the higher moments behave quite differently from what classical diffusion predicts. In particular, the autocorrelation function does not decay to zero, but instead saturates to a finite, size-independent value on the order of $5 \%$. To understand this effect, we explore more directly the consequences of the mapping between the monopole motion and quantum hopping on a Cayley tree. 

Ignoring weak correlations due to the spin ice background (see in App.~\ref{app:trees}), one can regard each bond on the tree (of coordination $3$) as having a $1/3$ probability of being cut, because the corresponding spin is static. This immediately implies, for instance, that in $1/27$ of the initial sites a monopole has no available moves, and so on. Classical percolation on a Cayley tree was solved in Ref.~\onlinecite{essam}; the percolation threshold is when half the links are cut, so constrained spin ice is well above this classical percolation threshold. 
Nevertheless, the \emph{quantum} problem is not trivially delocalized~\cite{harris1982}. Extended states appear in the quantum problem when a fraction $\alt 0.4$ of bonds are cut (i.e., for $p \agt 0.6$ of having a bond). As this is close to $2/3$ one might expect some signatures of proximity to percolation to appear in the properties of the spectra and wavefunctions. This is indeed what we see. 
%
%

\section{\label{sec:dos}
Density of states and wavefunction properties
             }
\begin{figure}[tb]
\begin{center}
\includegraphics[width = 0.48\textwidth]{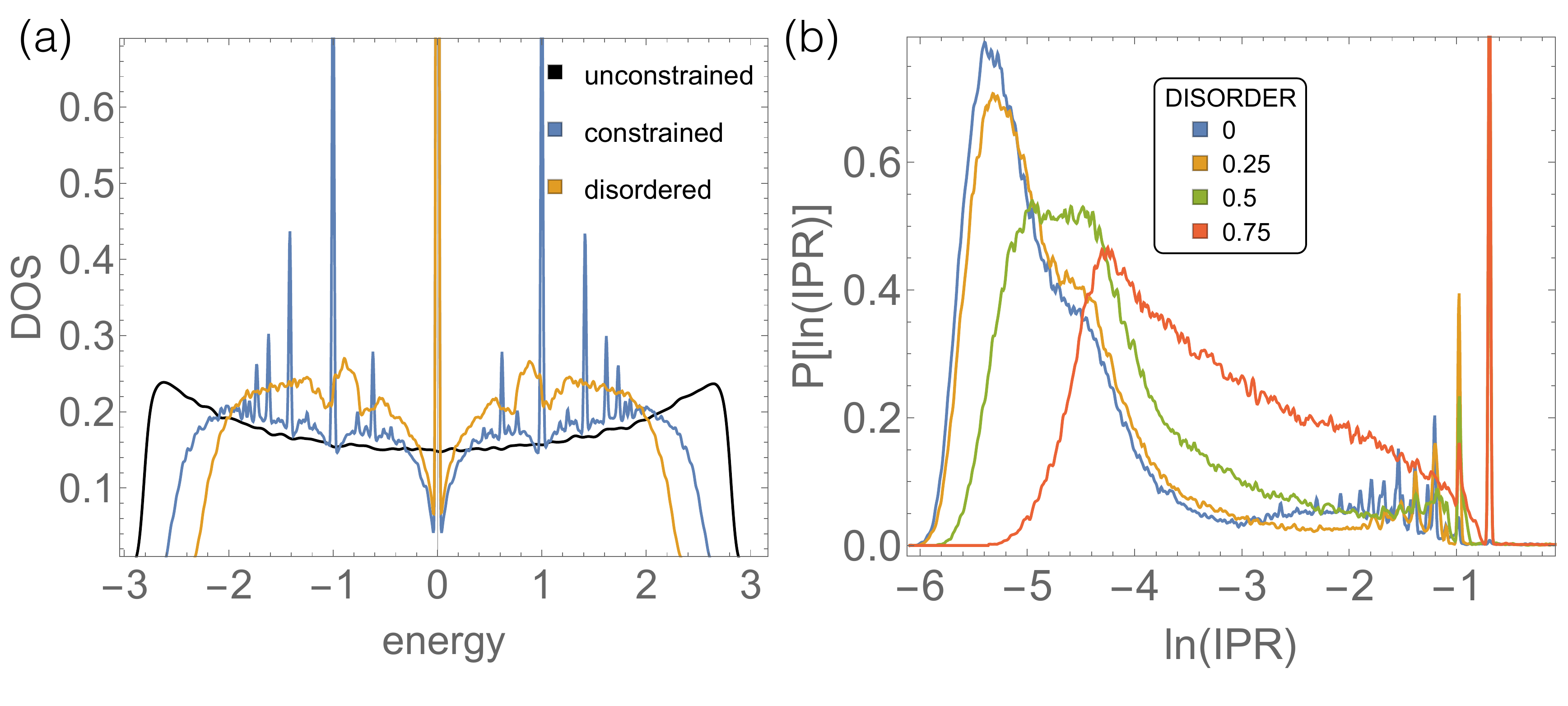}
\caption{(a)~One-monopole density of states vs. energy, for unconstrained, constrained, and constrained + disordered spin ice with disorder of $25 \%$ in the hoppings. The constraint creates disconnected clusters, which have discrete widely spaced levels. The disconnected clusters are all identical (in the absence of disorder in the hopping) so their levels coincide, giving sharp peaks in the DOS. Even relatively weak disorder smears out these peaks. (b)~Distribution of inverse participation ratios for the constrained case with various values of hopping disorder. The many sharp peaks corresponding to small clusters are smeared by disorder. However, the peak at the inverse participation ratio (IPR) $\mathcal{I}_2 = 1/2$ is unaffected by disorder, since the eigenstates on two-site clusters are the same no matter what the hopping amplitude is.}
\label{rrg1}
\end{center}
\end{figure}

We first consider the density of states. In the absence of a constraint, this is a regular function of energy; constraints give rise to peaks in the DOS corresponding to short disconnected clusters (see Fig.~\ref{rrg1}). 
Disorder in the hopping smears these peaks out, since the peak energies corresponding to different clusters are at different energies in the disordered system. There are also corresponding peaks in the inverse participation ratio (IPR), which is defined as $\mathcal{I}_2 \equiv \sum\nolimits_i |\psi(i)|^4$ and counts the number of states on which a wavefunction has appreciable support. In the absence of constraints, essentially all wavefunctions are delocalized over the entire tree, and the monopole is ergodic. The constraint drastically changes this: in addition to the states localized on classically disconnected clusters, which give discrete, evenly spaced peaks in the IPR distribution, there is a background of quantum localized states that live on one-dimensional segments of the percolation cluster, and experience standard quantum localization (as a result of the random self-energies due to small side chains hanging off the cluster~\cite{harris1982}). Localized and delocalized states are interspersed in the same energy window, without any apparent pattern (see App.~\ref{app:wavefunction}). 

\begin{figure}[tb]
\begin{center}
\includegraphics[width = 0.48\textwidth]{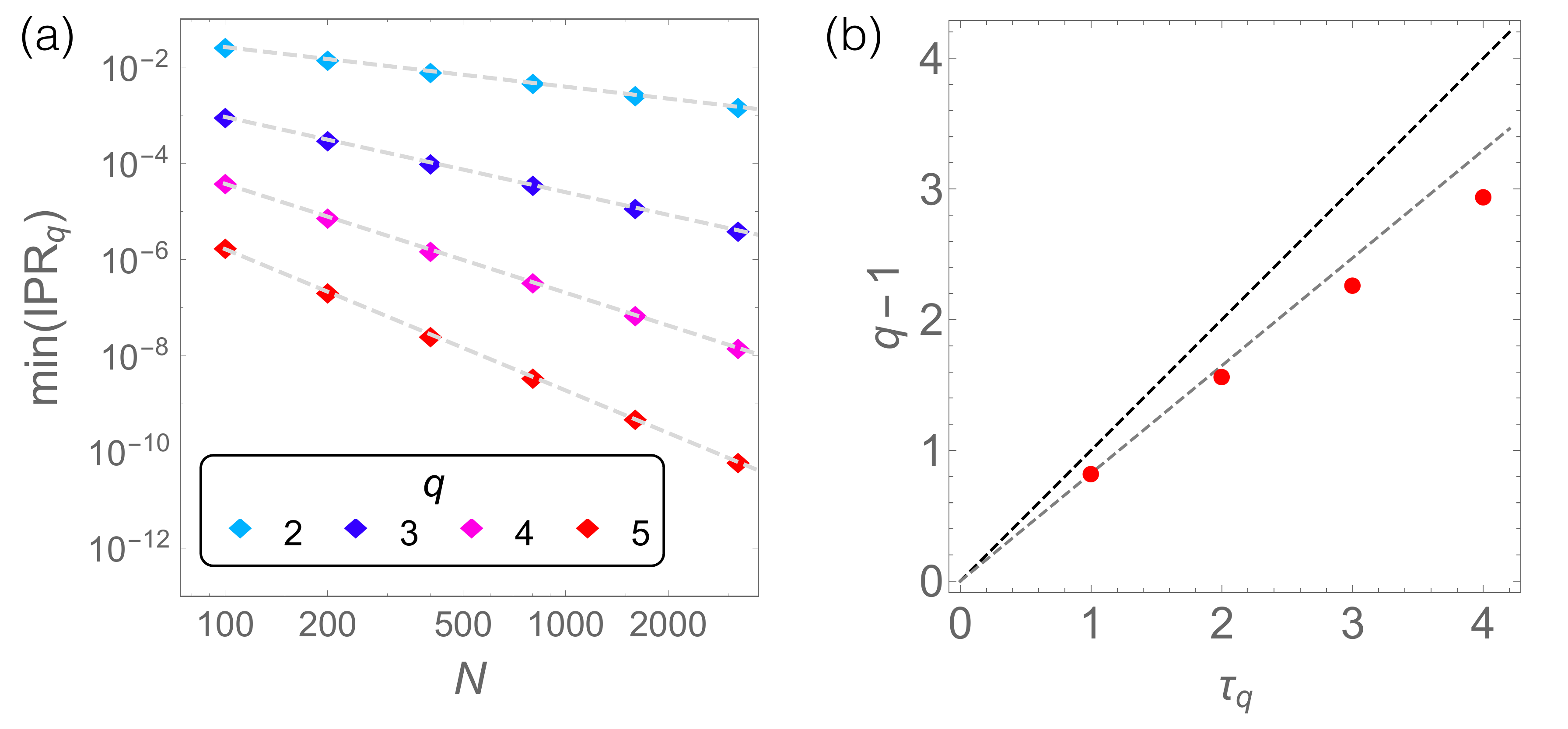}
\caption{(a)~Scaling of the generalized IPR $\mathcal{I}_q$ of the least localized state (defined in the main text) vs. tree size. All IPRs decrease algebraically with tree size, as $\mathcal{I}_q \sim N^{-\tau_q(q - 1)}$, where $\tau_q$ is plotted vs. $q - 1$ in panel (b). Ergodic states should have $\tau_q = 1$ (black line). The least localized state is therefore non-ergodic; the slight downward curvature of $\tau_q$ (relative to the gray line) indicates that these states are weakly multifractal.}
\label{rrg2}
\end{center}
\end{figure}

To investigate whether the delocalized states are ergodic or critical, we now explore the properties of the \emph{least} localized state in a typical sample, as a function of system size. The rationale for this choice is to understand whether \emph{any} ergodic states exist. We find (Fig.~\ref{rrg2}) that the IPR of the least localized state scales as $N^{-0.82}$ (as opposed to $N^{-1}$ as one would expect for an ergodic sample). 
Thus even the most delocalized states are non-ergodic. 

We explore the structure of these states further by computing their generalized IPRs, $\mathcal{I}_q \equiv \sum\nolimits_i |\psi(i)|^{2q}$. These go as $q$-dependent power laws, $\mathcal{I}_q \sim N^{-\tau_q}$. The anomalous exponent $\tau_q$ is plotted in Fig.~\ref{rrg2}(b): if the wavefunctions were fractal but otherwise structureless, we would see $\tau_q \propto q - 1$. The curve bends slightly downward, indicating that the states are (weakly) multifractal. This multifractal behavior is explored in more detail in App.~\ref{app:wavefunction}. The finding of non-ergodic monopole states is interesting because much attention has gone into exploring such states in the context of many-body localization. Perhaps surprisingly, the absence of \emph{any} ergodic wavefunctions appears to be compatible with quantum diffusion, as our results here indicate. 
%
%

\section{\label{sec:conductivity}
Conductivity
             }
\begin{figure*}[htbp]
\begin{center}
\includegraphics[width = 0.85 \textwidth]{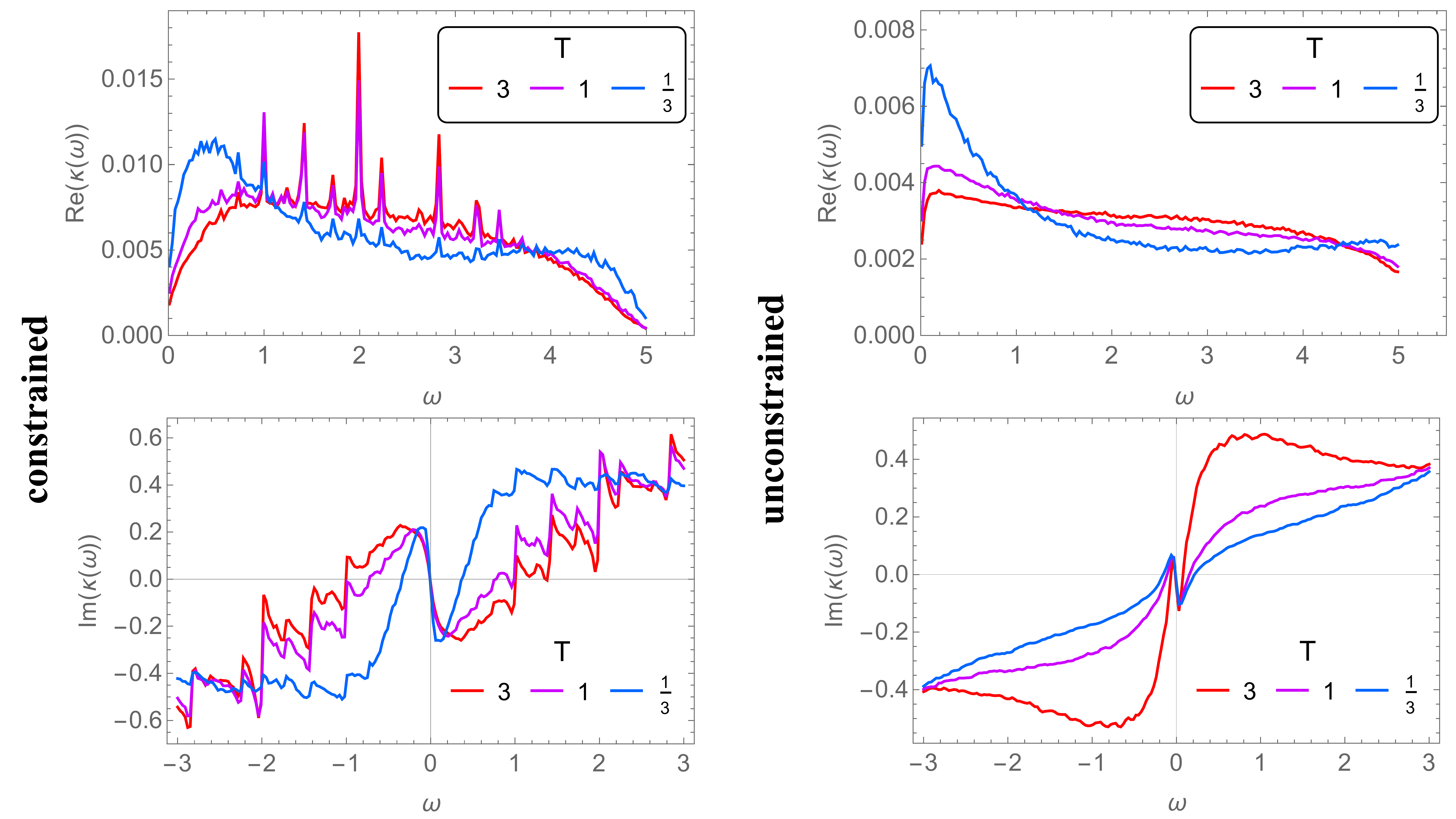}
\caption{Real and imaginary parts of the monopole conductivity $\kappa(\omega)$ at high, intermediate, and low temperatures for the constrained (left) and unconstrained RRGs. All curves are fixed to have the same normalization, so that they can be shown on the same scale.}
\label{kappatemp}
\end{center}
\end{figure*}

We finally briefly remark on the dynamical spin structure factor $S_{\alpha\beta}(q,\omega)$ and other dynamical response functions; we choose $\alpha\beta = xx$ for concreteness. 
$S_{xx}(q,\omega)$ is proportional to the Fourier transform of the autocorrelation function of the total spin; this is, in turn, proportional to the density of monopoles times the displacement of each monopole at time $t$. Assuming independently diffusing monopoles, $S_{xx}(q,\omega) \sim Dq^2/(D^2 q^4 + \omega^2)$, so the monopole conductivity $\kappa_{xx}(\omega) \sim D \omega^2 / (D^2 q^4 + \omega^2)$. In the d.c. limit, we expect this formula to be invalid, because collective effects become important. In App.~\ref{app:conductivity}, we explore the frequency- and temperature-dependence of $\kappa(\omega)$, using numerical studies of random regular graphs (RRGs) to mitigate finite-size effects (see Fig.~\ref{kappatemp}). 
For the unconstrained model this analysis gives diffusion, with a diffusion constant that is in good agreement with the directly determined one. For the constrained model, the ``apparent'' d.c. conductivity---defined by time-integrating the current-current correlator---seems to decrease as a function of observation time, even on the timescales where real-time dynamics sees diffusion. Note that $\kappa(\omega)$ is more sensitive to subleading non-analytic terms than the wavepacket width; thus, in regimes with coexisting localized and diffusive states, these quantities might disagree. 

In addition to this diffusive structure, response functions will generically have sharp peaks at predictable frequencies in the constrained case. However, these peaks are unstable to being smeared out by disorder, and therefore might not be visible in experiments involving non-Kramers ions, where structural disorder has been argued to act as a disordered transverse field to the spins~\cite{Savary2017dis,Benton2018}. 
Finally, our Kubo calculations do yield a reactive part in the conductivity that changes sign as a function of frequency (see App.~\ref{app:conductivity}); this behaviour -- which emerges here from a microscopic model -- is reminiscent of the one observed in Ref.~\onlinecite{pan2016}, and was explained phenomenologically by attributing a band mass to the monopoles~\cite{Armitage2018}. 
%
%

\section{\label{sec:discussion}
Discussion
             }
In this work we discussed the dynamics of an isolated quantum monopole in spin ice, in the temperature regime where the system samples the entire classical ground-state manifold. A mapping to quantum percolation on a Cayley tree exists; a finite fraction of monopole states are localized, while the majority are delocalized yet non-ergodic in Fock space. Despite these anomalous properties, transport is diffusive, with a well-defined diffusion constant that depends only quantitatively on dynamical constraints or extrinsic disorder. An important question for future work is to identify observable consequences of these anomalous and localized states~\cite{Hart2020}, e.g., in nonlinear response. It would also be interesting to see how much of this phenomenology survives at still lower temperatures, e.g., when the monopole lives on top of a true quantum spin-liquid ground state rather than a classical mixture. 
Although we considered square ice as a proxy for pyrochlore ice, our central results hinge on a Cayley tree mapping that applies equally well in both cases; thus, we expect these conclusions to extend to pyrochlore ice. 
An important implication of this work is that QSI is a model experimental system for studying quantum percolation, and related localization phenomena, on Cayley trees. In this context, it is interesting that multifractal wavefunctions appear to coexist with diffusive transport. 

Our results are relevant not only in searching for QSL behaviour in QSI candidate materials, but also in the context of implementing QSI (and QSLs in general) in quantum annealers -- a possibility that has come to the fore recently~\cite{King2021}. Indeed, quantum annealer implementations operate precisely in the temperature regime of the classical incoherent spin background discussed in our work. Ref.~\onlinecite{King2021} demonstrates that, while quantum square ice orders at zero temperature and the excitations are log confined, finite temperature suppresses the ordered phase and allows the excitations to wonder unimpeded across the lattice (over the relevant time and length scales). This opens the door to implementing and investigating quantum spin ice and quantum spin liquids in quantum annealers, by exploiting precisely the semiclassical finite-temperature regime that we investigate here; moreover, implementing QSLs in quantum annealers paves the way to other types of gauge theories, the toric code and colour and surface codes that may put quantum annealers back on the map towards topological quantum information processing and quantum computing. 
%
%

\section*{Acknowledgements}
We thank N.P. Armitage for helpful discussions. This work was supported in part by the Engineering and Physical Sciences Research Council (EPSRC) Grants No.~EP/K028960/1,~EP/M007065/1, and~EP/P034616/1 (CC), by the NSF Grant DMR-1653271 (SG), and by the Deutsche Forschungsgemeinschaft under grant SFB 1143 (project-id 247310070) and the cluster of excellence ct.qmat (EXC 2147, project-id 390858490) (R.M.). 
%
%
\appendix

\section{\label{app:trees}
From quantum spin ice to trees
             }
In this Appendix we describe the mapping between quantum spin ice clusters and random regular graphs, compare the two cases numerically, and then describe ``transport'' properties on the random regular graph. 

\subsection{Geometric considerations}

\begin{figure*}[htbp]
\begin{center}
\includegraphics[width = 0.8\textwidth]{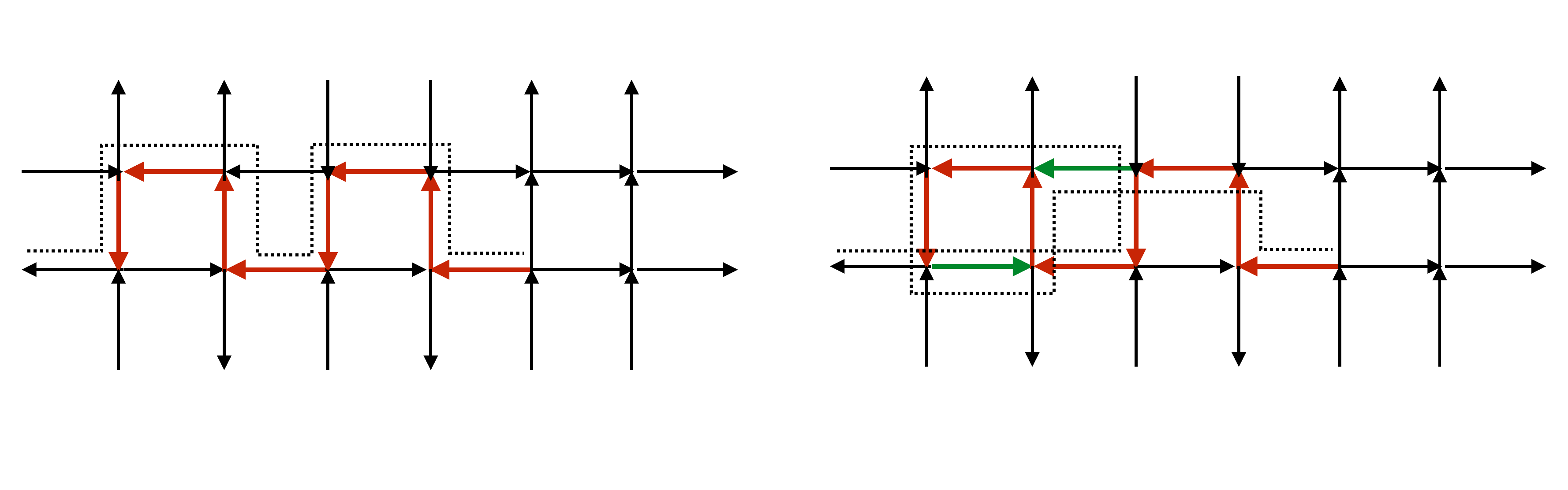}
\caption{Two inequivalent monopole trajectories that both lead to the same set of flipped spins (dark red). The second trajectory involves flipping some spins (dark green) twice, and thus has longer path length.}
\label{minpairs}
\end{center}
\end{figure*}

We first discuss the mapping from the dynamics of quantum spin ice to that on a Cayley tree. The essential idea, as explained in the main text, is that the dynamics of spin ice is essentially treelike: different trajectories from $A$ to $B$ leave distinct trails of flipped spins behind, and therefore cannot interfere. Thus, as outlined in the main text, one can approximate the spin ice dynamics using a Cayley tree of the appropriate random connectivity ($\sim \frac{2}{3}$). This mapping leaves out some features: first, the orientations of the three flippable spins surrounding a monopole are not entirely independent, but could in principle be correlated via the ice rules; and second, there are specific pairs of paths that interfere, as shown in Fig.~\ref{minpairs}. This figure shows two paths that have the same trail and the same endpoints, but different path lengths and different phases. The minimal case with definite constructive interference consists of trajectories that go through two ``bends'' of the form shown. A trajectory that takes the first bend the long way and the second the short way will interfere constructively with one that takes the first bend the short way and the second the long way, yielding a correction. 

Neither effect is especially significant: the former reduces the fraction of unblocked paths to slightly below $\frac{2}{3}$, while the latter effect seems to require fine-tuned low-entropy pairs of paths. 

\begin{figure}[htbp]
\begin{center}
\includegraphics[width = 0.45\textwidth]{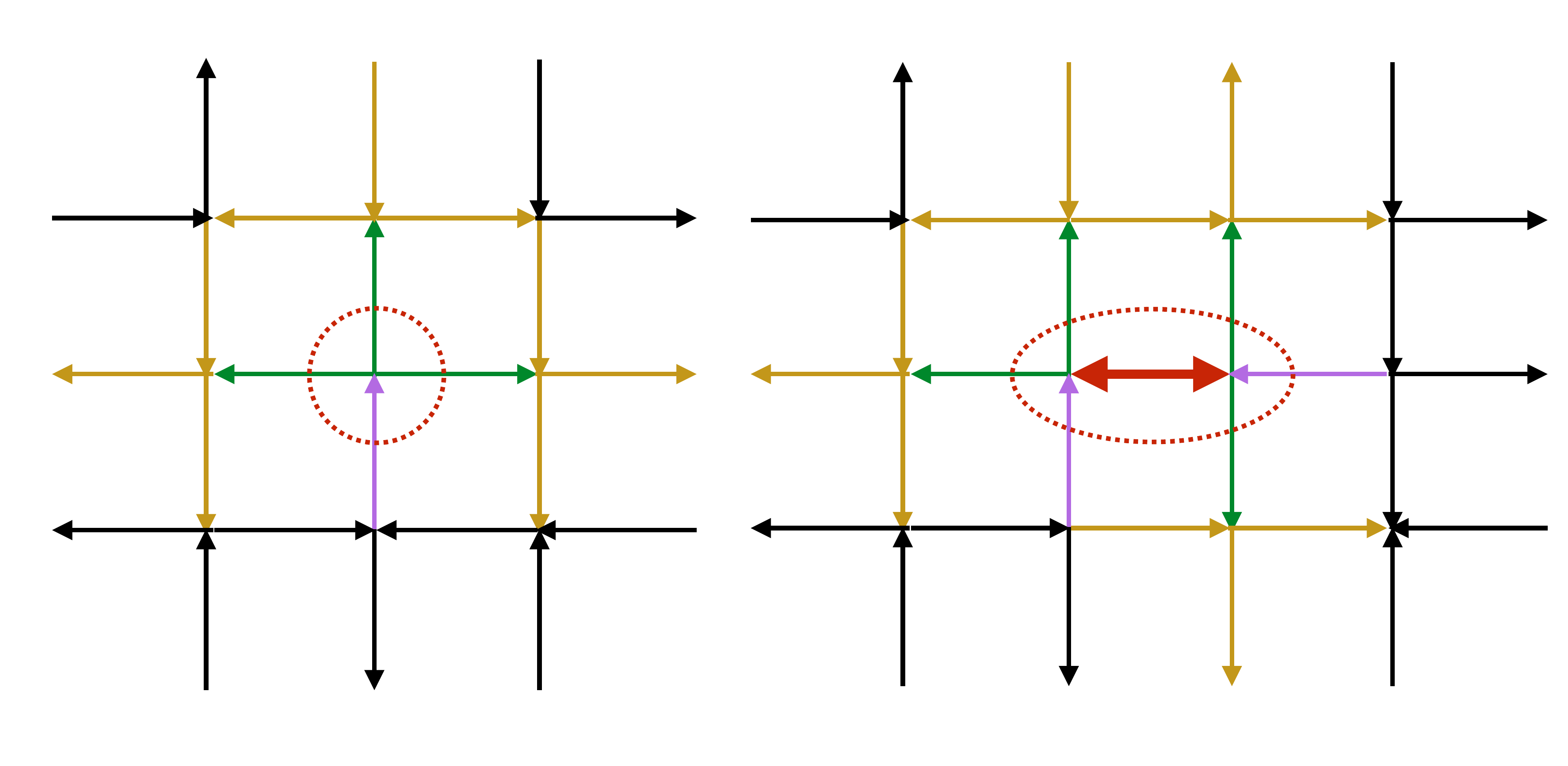}
\includegraphics[width = 0.35\textwidth]{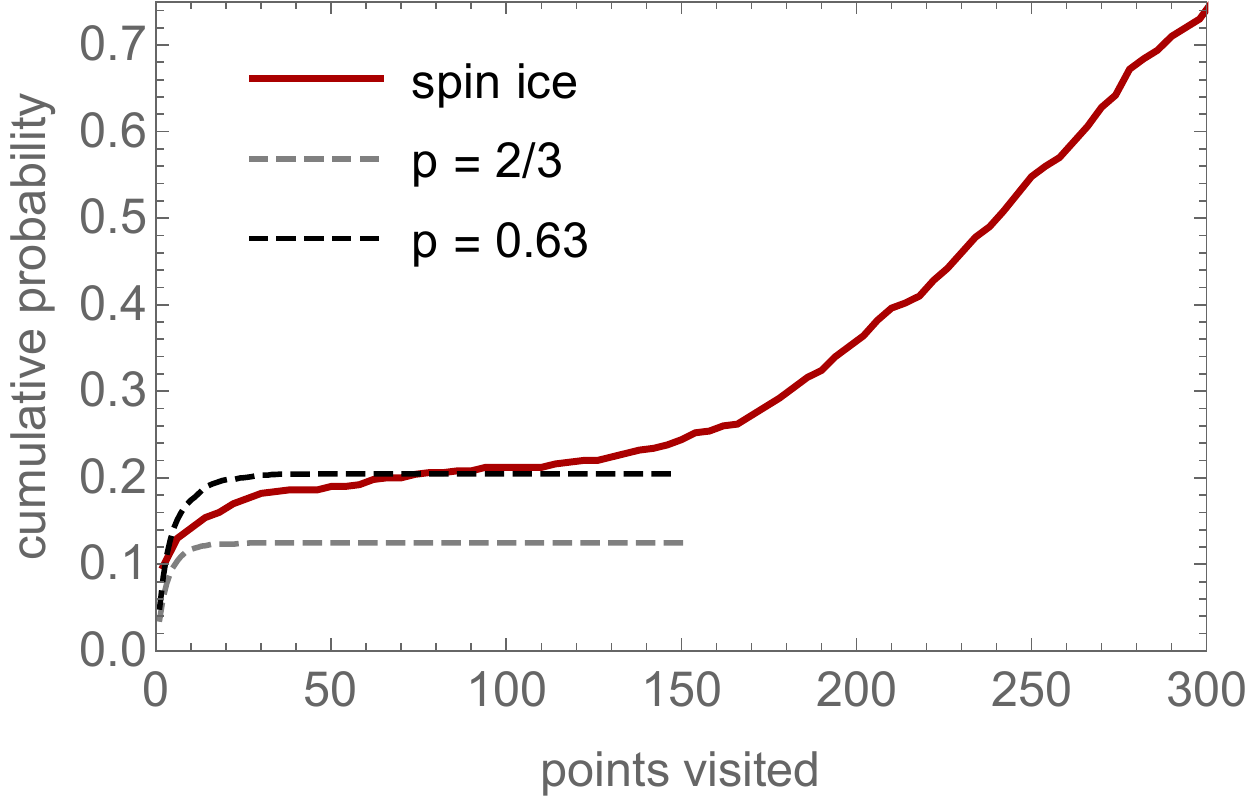}
\caption{Left/center: Configurations of path length zero and one. The monopole itself (and the region in which it can move) are indicated in red. Each monopole position has one energetically blocked spin, marked in purple. Potentially flippable spins (frozen because of the transverse field configuration) are marked in green, while spins that ``cage'' the monopole by immobilizing the green spins are themselves marked in brown. Right: Cluster sizes for square ice, compared with the results for random regular graphs with $p = 0.63$ and $p = \frac{2}{3}$ connectivity. All the models have a finite fraction of very small clusters; this fraction quantitatively matches for $p = 0.63$ (which is the appropriate connectivity in square ice, once one incorporates nearest neighbor correlations).}
\label{configs}
\end{center}
\end{figure}

In the tree approximation, $\frac{1}{27}$ monopoles are entirely blocked; $\frac{4}{243}$ rattle between two sites; etc. These two types of localized configurations are shown in Fig.~\ref{configs}; for a discussion of the distribution of small cluster sizes within this tree model we refer to Ref.~\onlinecite{essam}. The square ice model has further correlations imposed by the ice rule (so the states of the three edges around a monopole are not strictly independent). Analyzing these carefully gives us that in fact the fraction of flippable spins is $p \approx 0.626 < \frac{2}{3}$. By contrast, in pyrochlore spin ice, these correlations are unimportant so $p \approx \frac{2}{3}$ to good accuracy. Counting path distributions in the constrained spin ice model agrees quantitatively with this prediction (Fig.~\ref{configs}). 

\subsection{Numerical comparisons}

\begin{figure*}[htbp]
\begin{center}
\includegraphics[width = 0.95\textwidth]{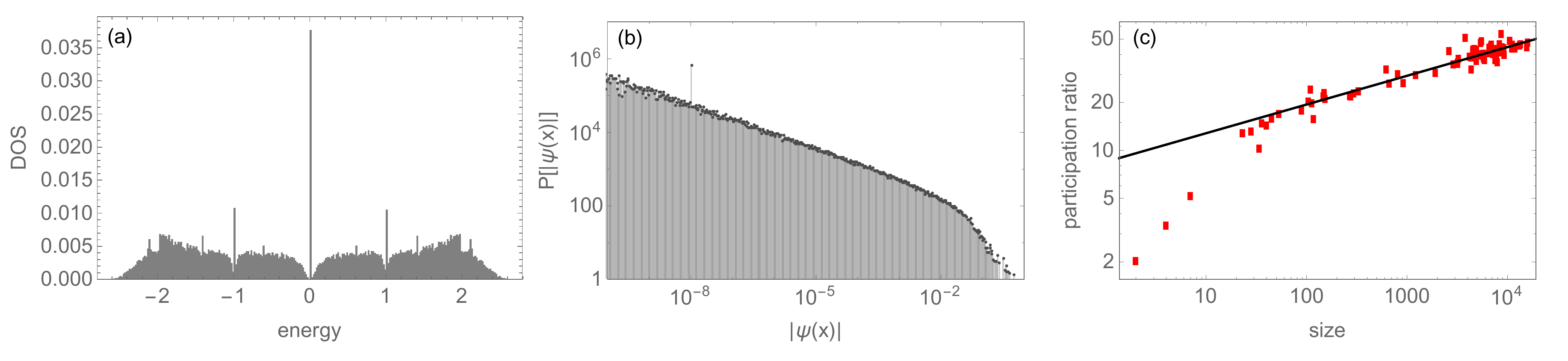}
\includegraphics[width = 0.95\textwidth]{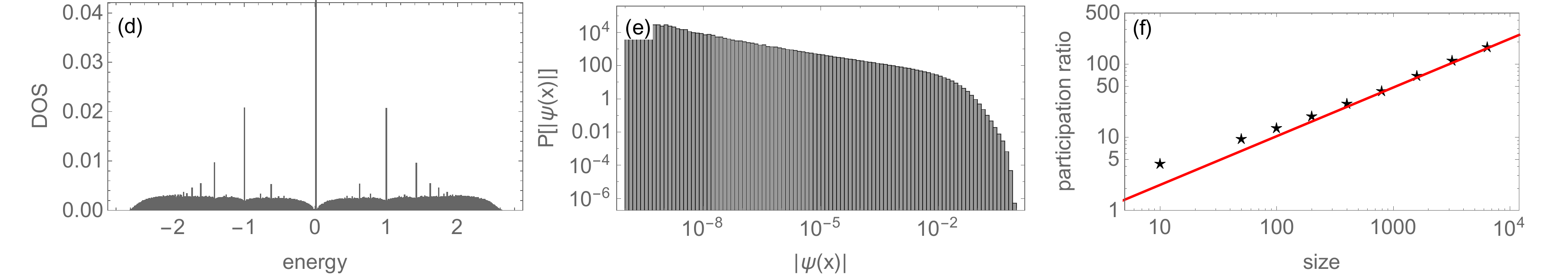}
\caption{Comparison of exact diagonalization data on microscopic trees (top) and RRGs (bottom). (a)~Density of states of monopoles on constrained square ice, showing peaks due to disconnected clusters and a pronounced depletion near $E = 0$. (b)~Probability density of $|\psi(x)|$; at small $x$ this goes as $P(|\psi|) \sim |\psi|^{-0.8}$. (c)~Typical participation ratio of wavefunctions on a cluster vs. size of cluster: $\mathrm{IPR}_{typ.}(L) \sim L^{-0.18}$. Lower panel: corresponding results on random regular graphs. For this case one has $P(|\psi|) \sim |\psi|^{-0.36}$ [panel~(e)] and $\mathrm{IPR}_{typ.}(L) \sim L^{-0.66}$ [panel~(f)]. }
\label{compfig}
\end{center}
\end{figure*}

We now turn to a comparison of numerical calculations on the spin ice model within the tree approximation (we will call this the microscopic tree in what follows) and those on random regular graphs with the same average connectivity. The algorithm for constructing and analyzing trees is outlined in the main text: briefly, one begins with a randomly generated spin ice configuration, flips a spin to create a monopole-antimonopole pair, moves the antimonopole to the edge of the system, and then constructs a graph of paths that the monopole is allowed to traverse. This graph is treelike, by construction, and we call it the microscopic tree. The microscopic trees and the random regular graphs have somewhat different drawbacks. On the one hand, the microscopic tree incorporates local correlations that are neglected by random regular graphs (RRGs), and thus gives a more accurate picture of the local physics. On the other hand, the microscopic trees \emph{terminate} when they hit the edge of the sample, and a finite fraction of the nodes of the sample are at the edge. The states near the edge generally have lower coordination and are likelier to form disconnected clusters (or localized wavefunctions on connected clusters). The RRG avoids these spurious edge effects by terminating the tree with large loops rather than cut edges. Comparing the behavior of the two models, the RRG consistently seems less localized than the microscopic tree (Fig.~\ref{compfig}). 

As Fig.~\ref{compfig} shows, despite these quantitative differences, both models show the main features we are interested in: for example, both models have wavefunctions that appear multifractal at these scales, filling in only a small fraction of the classical clusters they live on, although the multifractal exponents are different. Likewise, the density of states has regularly spaced peaks, as discussed in the main text, although the peak heights are suppressed. We note the suppression of density of states at energies near the peaks; this feature is robust, but we do not have an analytic understanding of it at present. 
%
%

\section{\label{app:wavefunction}
Wavefunction properties
            }
For the temporal dynamics shown in the main text, we were able to work with the microscopic trees and avoid the edge effects because a monopole initialized at the root of the tree could not reach the edge on the studied timescales. Unfortunately this is no longer the case for wavefunction properties or optical conductivity, so we work with RRGs. For the wavefunction properties, which we consider first, we expect the RRG model to be quantitatively correct. For optical conductivity, which we discuss next, turning to RRGs entails an additional approximation, but we still expect that the qualitative behavior is correctly captured. 

\begin{figure}[htbp]
\begin{center}
\includegraphics[width = 0.36\textwidth]{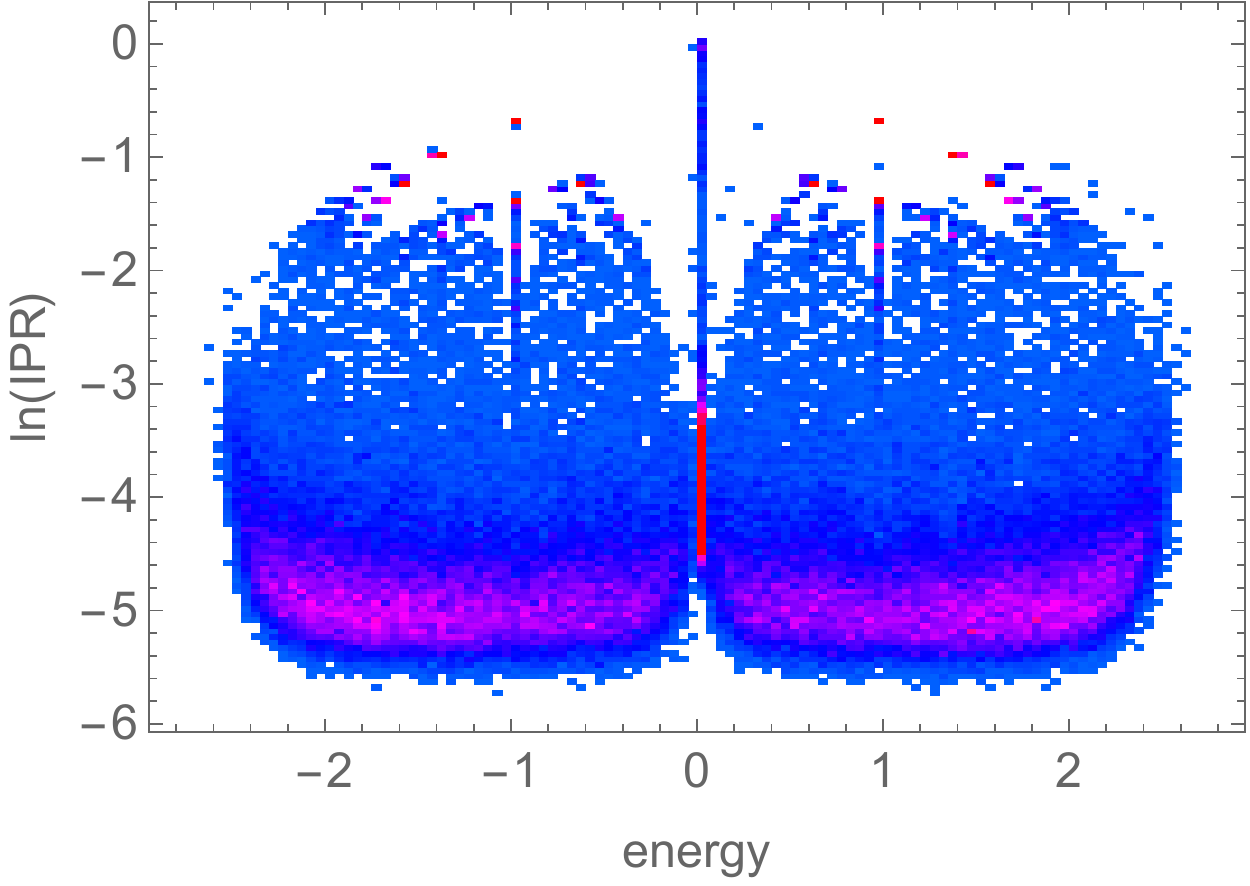}
\includegraphics[width = 0.375\textwidth]{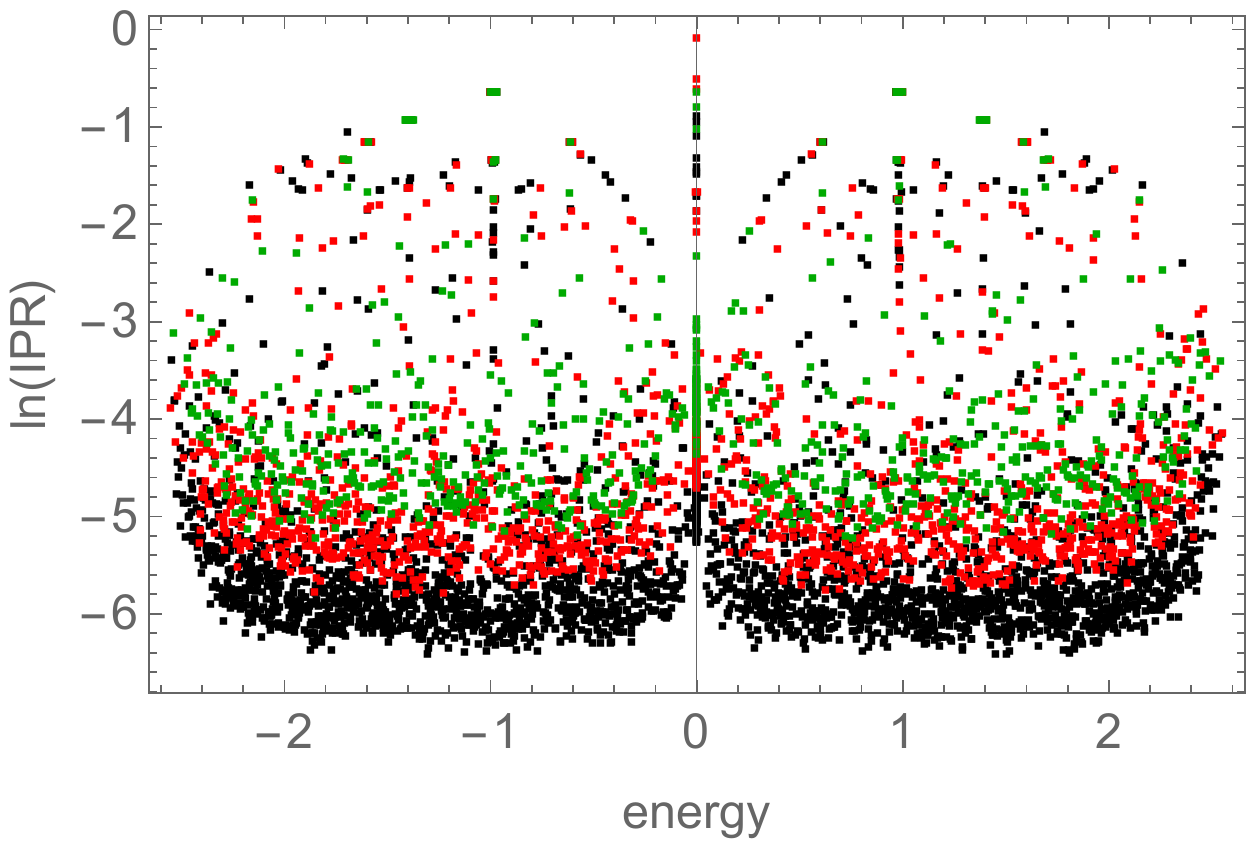}
\includegraphics[width = 0.375\textwidth]{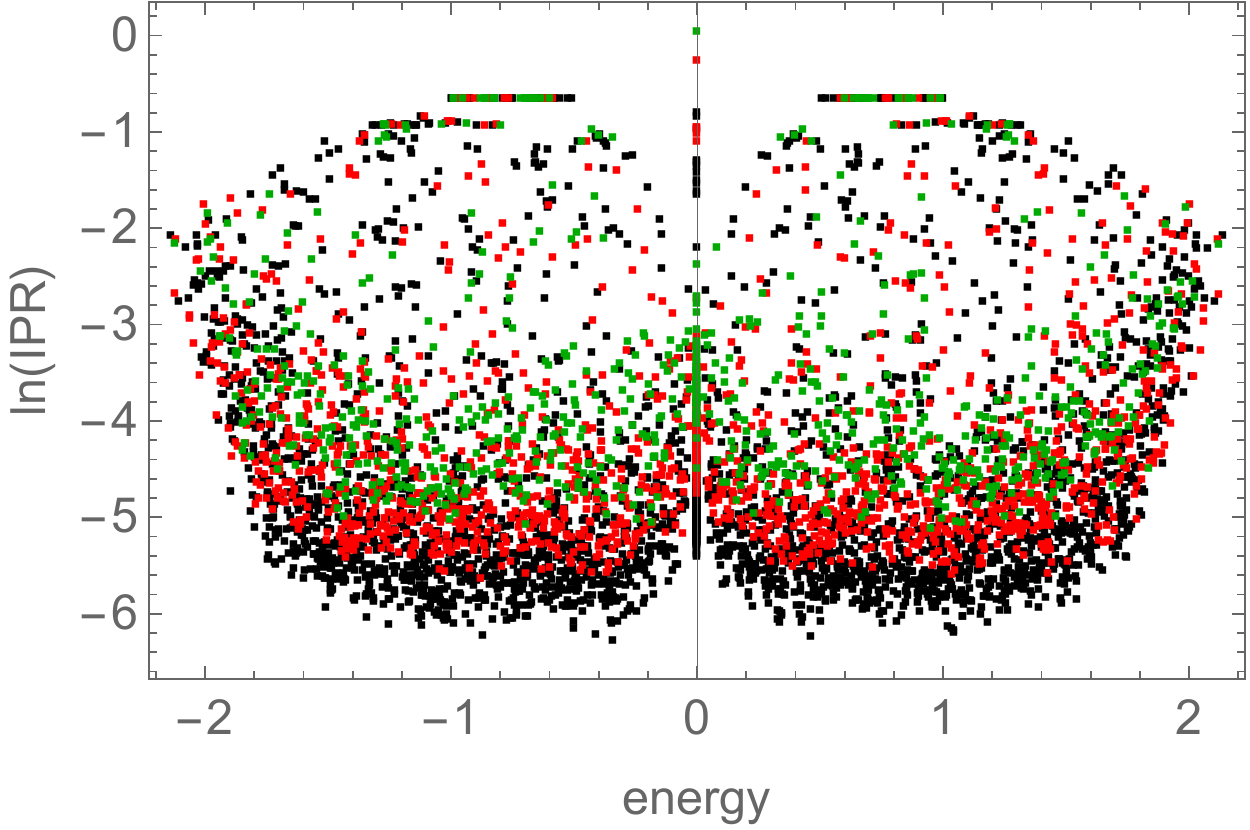}
\caption{Behavior of inverse participation ratio (IPR) vs. energy. Left: density of states as a function of IPR and energy in the constrained model; the color scheme goes from blue (low density) to red (high density). Center: scatter plot of IPR and energy for three different system sizes: $L = 800, 1600, 3200$ (green, red, black). Right: same axes and color scheme, but with $33\%$ disorder in the hopping.}
\label{iprdata}
\end{center}
\end{figure}

Fig.~\ref{iprdata} plots the inverse participation ratio (IPR) vs. energy. To get meaningful numerical data, one must break the large degeneracies that exist in the clean constrained system; we do so by adding disorder of $1\%$ in the hoppings. In both the ``nearly clean'' and strongly disordered cases, wavefunctions are heterogeneous (with some much more localized than others) but there is no clear sign of a mobility edge. Rather, the wavefunctions at a given energy have a broad distribution of localization lengths. This is a distinctive feature of percolation on trees. At the sizes we have studied, none of the states appears to be fully delocalized. Moreover, in the clean case, one sees a sequence of isolated states that live on disconnected clusters, for which the IPR and energy are geometrically determined. These localized clusters are less distinct in the disordered case, but the two- and three-site clusters manifest themselves as horizontal lines at IPR $\frac{1}{2}, \frac{1}{3}$ respectively. 
%
%

\section{\label{app:conductivity}
Optical conductivity
             }
We now turn from wavefunction properties to dynamics. In particular, we will consider the frequency-dependent monopole conductivity $\kappa(\omega)$. In order to define this on the RRG we need to first map the structure of the monopole current operator onto the RRG. We begin with the microscopic tree, for which one can define a real-space monopole \emph{position} operator $\hat{X}$ that identifies the real-space position (along the direction $x$ for concreteness) of the monopole in each global spin configuration (i.e., node of the tree). This is the same operator that we used to compute real-time dynamics as described in the main text. The current is then $\hat{J}_x = \frac{d}{dt} \hat{X}$, from which one can derive the linear response optical conductivity via the Kubo formula as 

\beq
\mathrm{Re}[\kappa(\omega)] \propto \omega(1 - e^{-\beta\omega}) \int_{-\infty}^\infty dt\, e^{i\omega t} \, \langle \hat{X}(t) \hat{X}(0) \rangle. 
\eeq
where we are leaving out a factor of monopole density. The imaginary part of the conductivity follows from the Kramers-Kronig relation. 

To better understand the structure of the local current operator $\hat{\mathbf{J}}(\mathbf{r})$, we consider the equation of motion for the monopole density. First, we observe that $\hat{Q}(\mathbf{r}) = \sum_{\alpha} |\alpha\rangle \langle \alpha| \delta(\mathbf{r}_\alpha - \mathbf{r})$. By commuting $\hat{Q}(\mathbf{r})$ with the Hamiltonian and using the continuity equation, we conclude that the current across a link is given by

\begin{eqnarray}
\hat{J}_x(\mathbf{r}, \mathbf{r}') &=& i \sum\nolimits_{\alpha\beta} (|\alpha \rangle \langle \beta| - |\beta\rangle \langle \alpha|) 
\nonumber \\ 
&\times& 
\delta(\mathbf{r}_\alpha - \mathbf{r}) \delta(\mathbf{r}'_\beta - \mathbf{r}') (x_{\mathbf{r}'} - x_\mathbf{r}). 
\end{eqnarray}
Each pair of sites linked by the current operator can either have the same $x$ component or an $x$ component that is larger or smaller by one. 
The sites to which the monopole can hop from a given position are oriented essentially randomly because of the entropy of the ice manifold. Motivated by these observations, we define an approximate current matrix on an RRG as follows: any pair of connected configurations has an associated current operator that is $+i$ with probability $\frac{1}{4}$, $-i$ with probability $\frac{1}{4}$, and zero otherwise. This ignores certain local correlations (in the physical system one cannot, for example, have two links out of a site that both point rightward), but in practice this type of unphysical connectivity occurs at $2\%$ of nodes, so for simplicity we have not put in an extra rule to exclude it. 

\subsection{Temperature- and disorder-dependence}

We now explore the properties of $\kappa(\omega)$ in this RRG model. Fig.~\ref{kappatemp} in the main text plots out its temperature dependence. Note that this is temperature is set only by the kinetic energy of the monopoles: we are assuming throughout this work that the temperature is too high to allow the ground-state degeneracy to be lifted. In the constrained case, there is a clear sign-change in the imaginary part of the conductivity (recall that this is the \emph{reactive} response) that was previously experimentally observed and attributed to monopole inertia~\cite{Armitage2018}. This feature becomes more pronounced and moves out to higher frequencies at higher temperatures. Interestingly, the low-frequency conductivity increases as the temperature is decreased: the states near the bottom of the band seem to have a higher mobility than those near $E = 0$. 

\begin{figure}[htbp]
\begin{center}
\includegraphics[width = 0.4\textwidth]{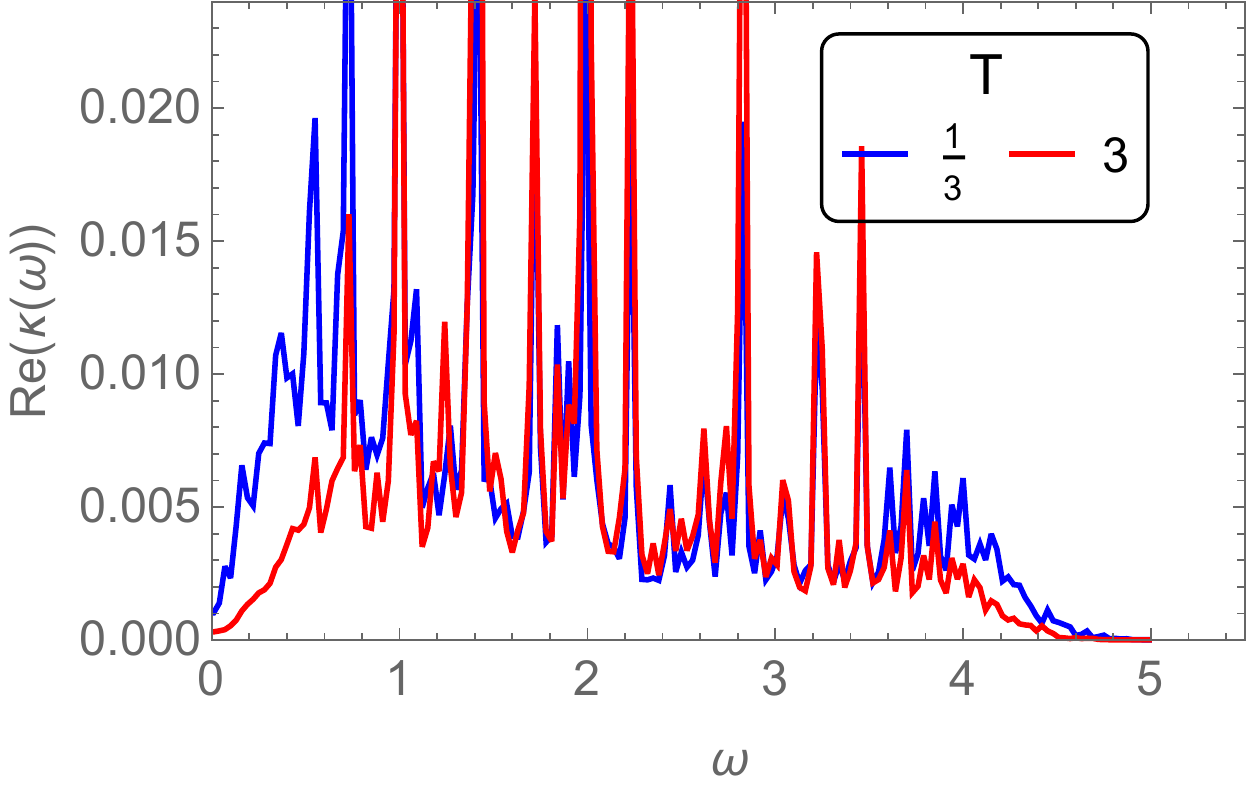}
\includegraphics[width = 0.4\textwidth]{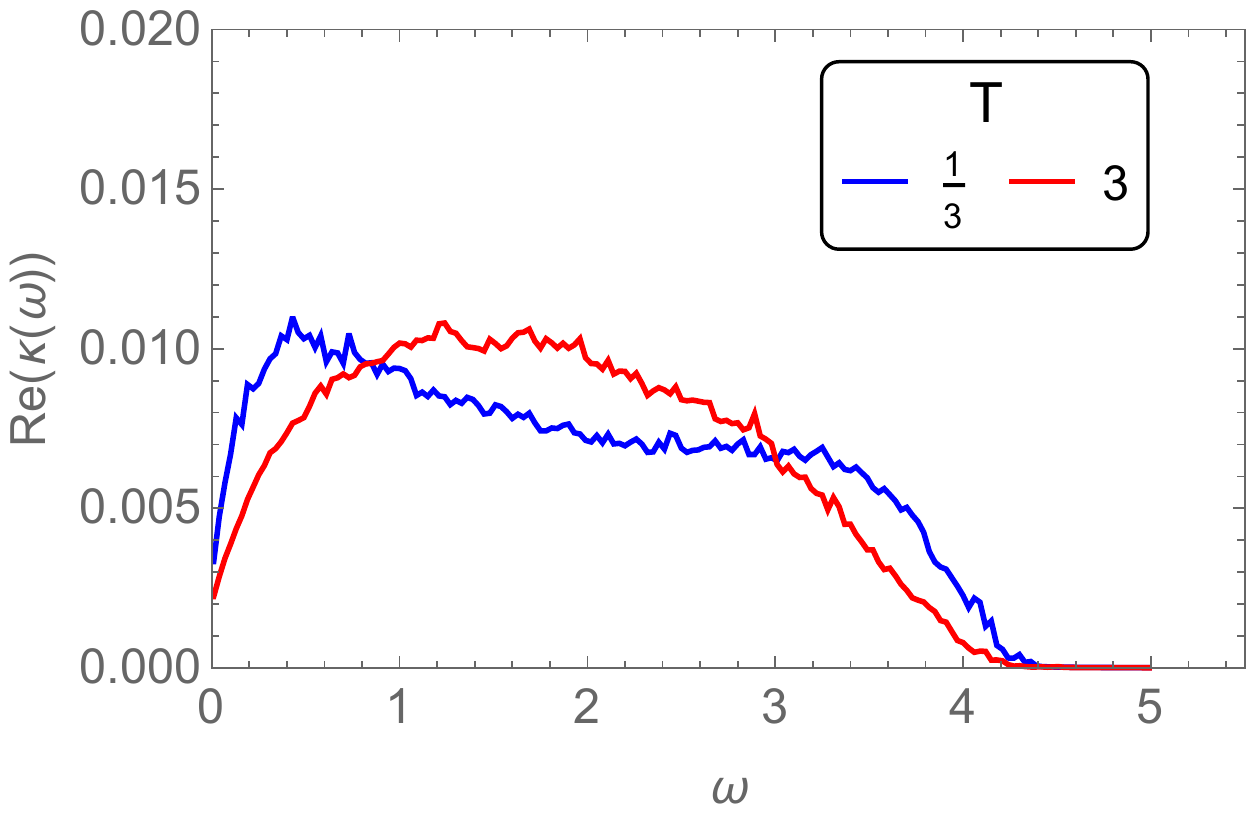}
\includegraphics[width = 0.4\textwidth]{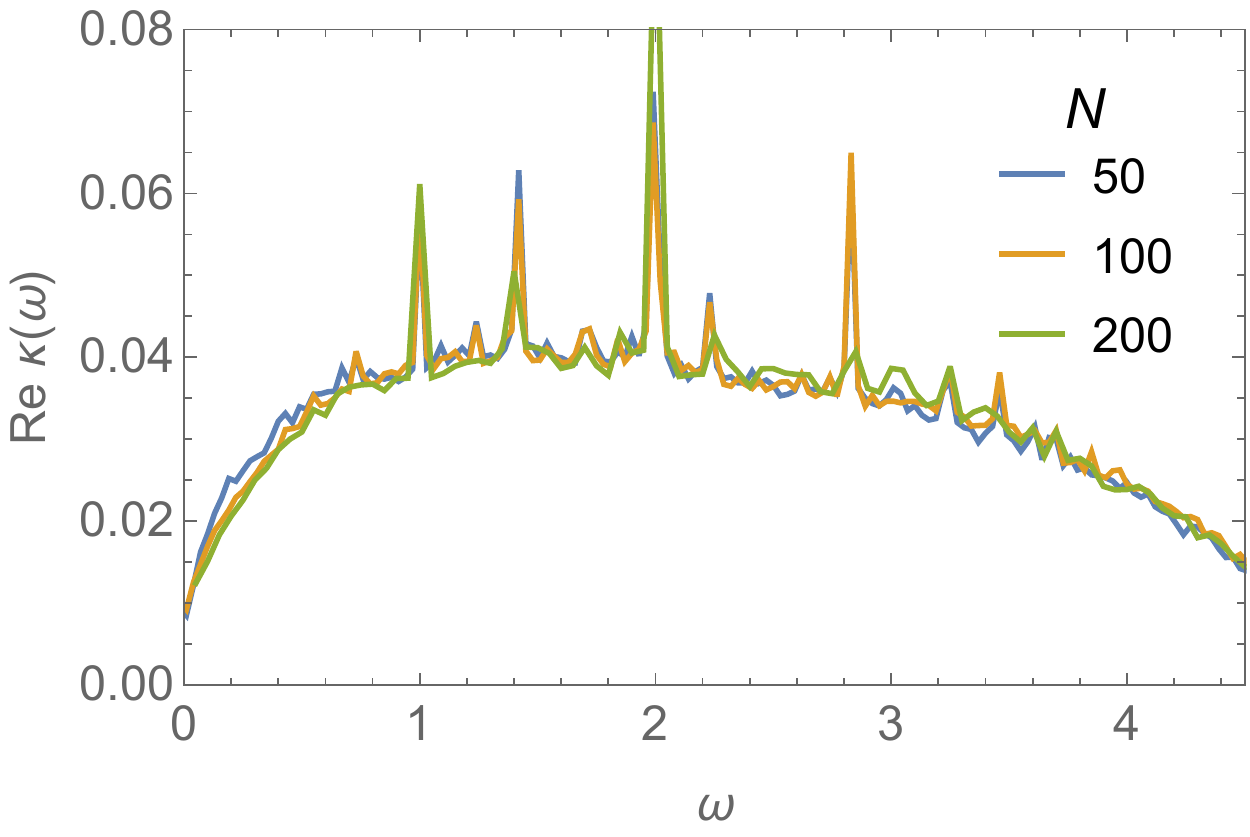}
\caption{High- and low-temperature conductivity for RRGs below percolation threshold [left] and with hopping drawn randomly from the interval $(0.5, 1)$ [center]. Right: size-dependence of conductivity at high temperatures in the constrained (but disorder-free) case.}
\label{morelocalized}
\end{center}
\end{figure}

We also briefly explore the behavior of $\kappa(\omega)$ in regimes where the hopping is disordered or the fraction of blocked directions is set artificially high (at $1/2$, i.e., percolation threshold for the RRG). Here, Fig.~\ref{morelocalized} shows the high- and low-temperature behavior in these cases. Below percolation threshold, the low-frequency behavior is much more clearly insulating (especially at high temperatures), and the peaks corresponding to small clusters are pronounced. Disorder smears the peaks, but also depletes low-frequency conductivity as one might expect. Finally, the rightmost panel of Fig.~\ref{morelocalized} shows the size-dependence of $\mathrm{Re} [\kappa(\omega)]$, which is very slight (although there is a slight drift toward having less weight at low frequencies). 

\subsection{Comparison with microscopic trees}

\begin{figure}[htbp]
\begin{center}
\includegraphics[width = 0.4\textwidth]{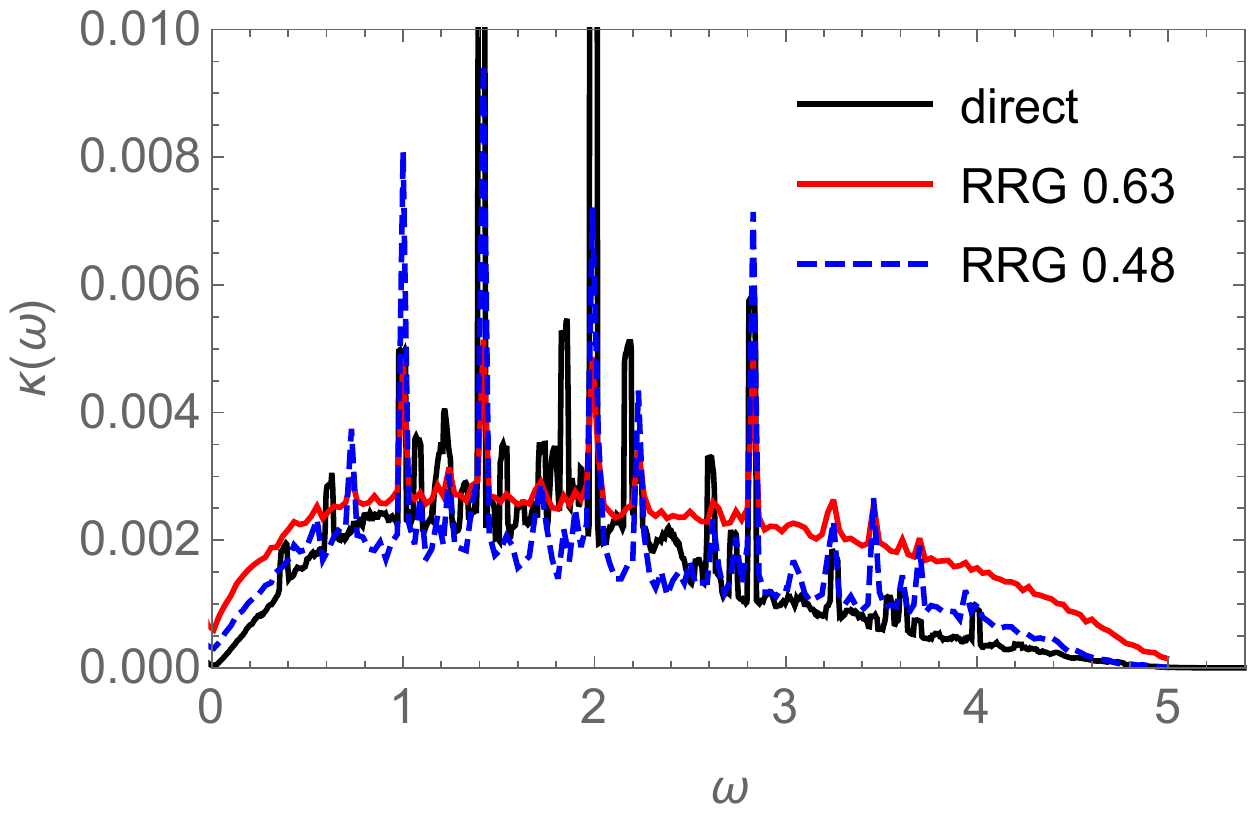}
\includegraphics[width = 0.4\textwidth]{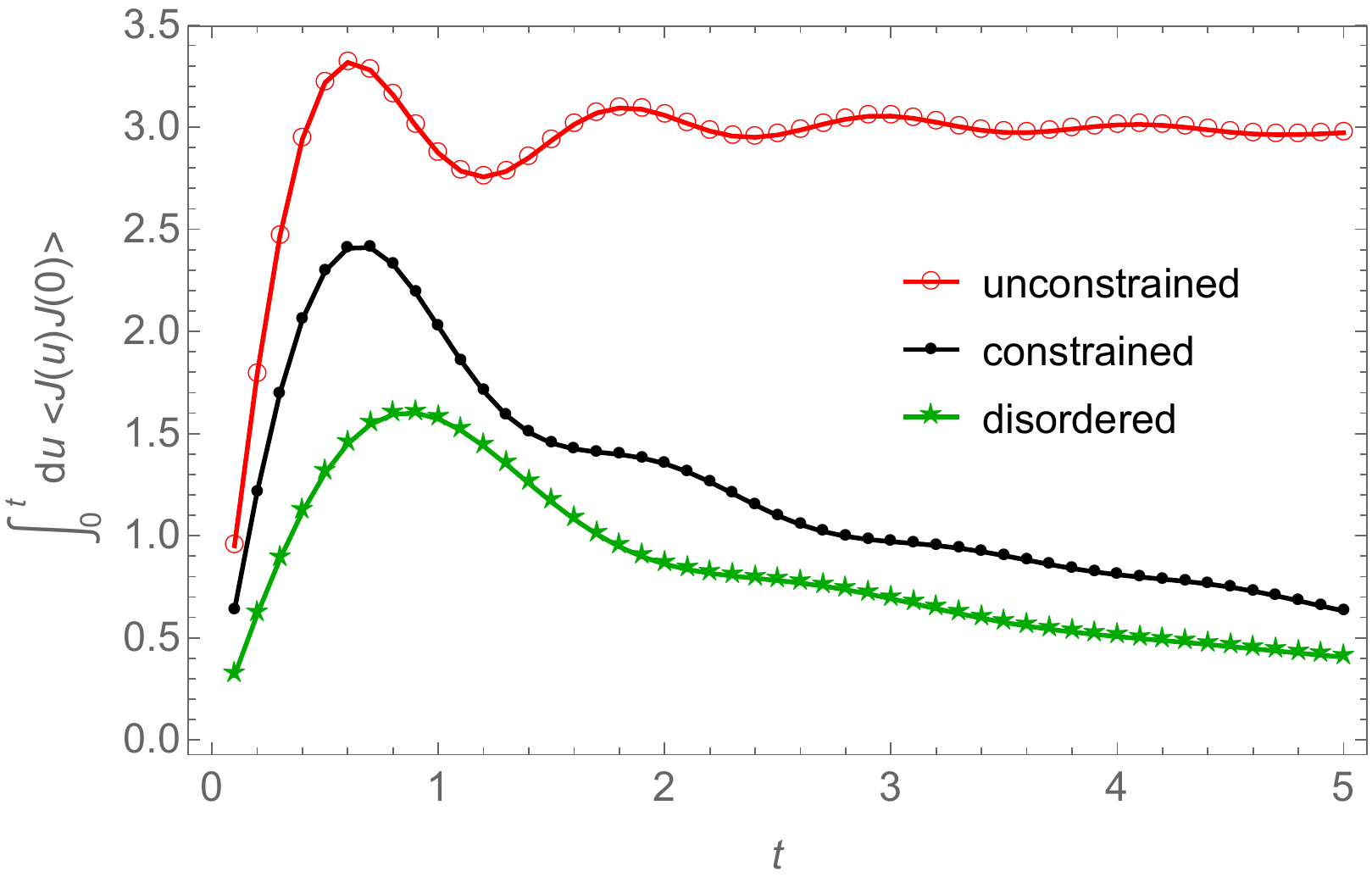}
\caption{Left: Real part of the conductivity from direct computation on the microscopic tree compared with results on the RRG at two different values of the connectivity. Edge effects seem to push the microscopic tree results to lower effective connectivity. Right: ``Apparent'' d.c. limit vs. time, from integrating the current-current correlator out to some time $t$. In the unconstrained case the model is cleanly diffusive, whereas the constrained and disordered cases have diffusion constants that drift downward with time.}
\label{2compfig}
\end{center}
\end{figure}

We now turn to a comparison between these results for the optical conductivity and results for the model with microscopic trees. In the microscopic tree model, we construct clusters of paths that are $20$ or fewer steps away from the root (where the monopole initially sits). As before, the adjacency matrix of this graph is the monopole Hamiltonian. If we compute the monopole conductivity using the Kubo formula, we find that the results are close to those for an RRG a little below the percolation threshold (Fig.~\ref{2compfig}). This effect seems to be due to anomalously localized states at the edge of the graph: these states have no weight on the initial monopole position, and thus do not affect the real-time dynamics studied in the main text, but do affect the conductivity. 

\subsection{Low-frequency limit}

\begin{figure*}[htbp]
\begin{center}
\includegraphics[width = 0.66 \textwidth]{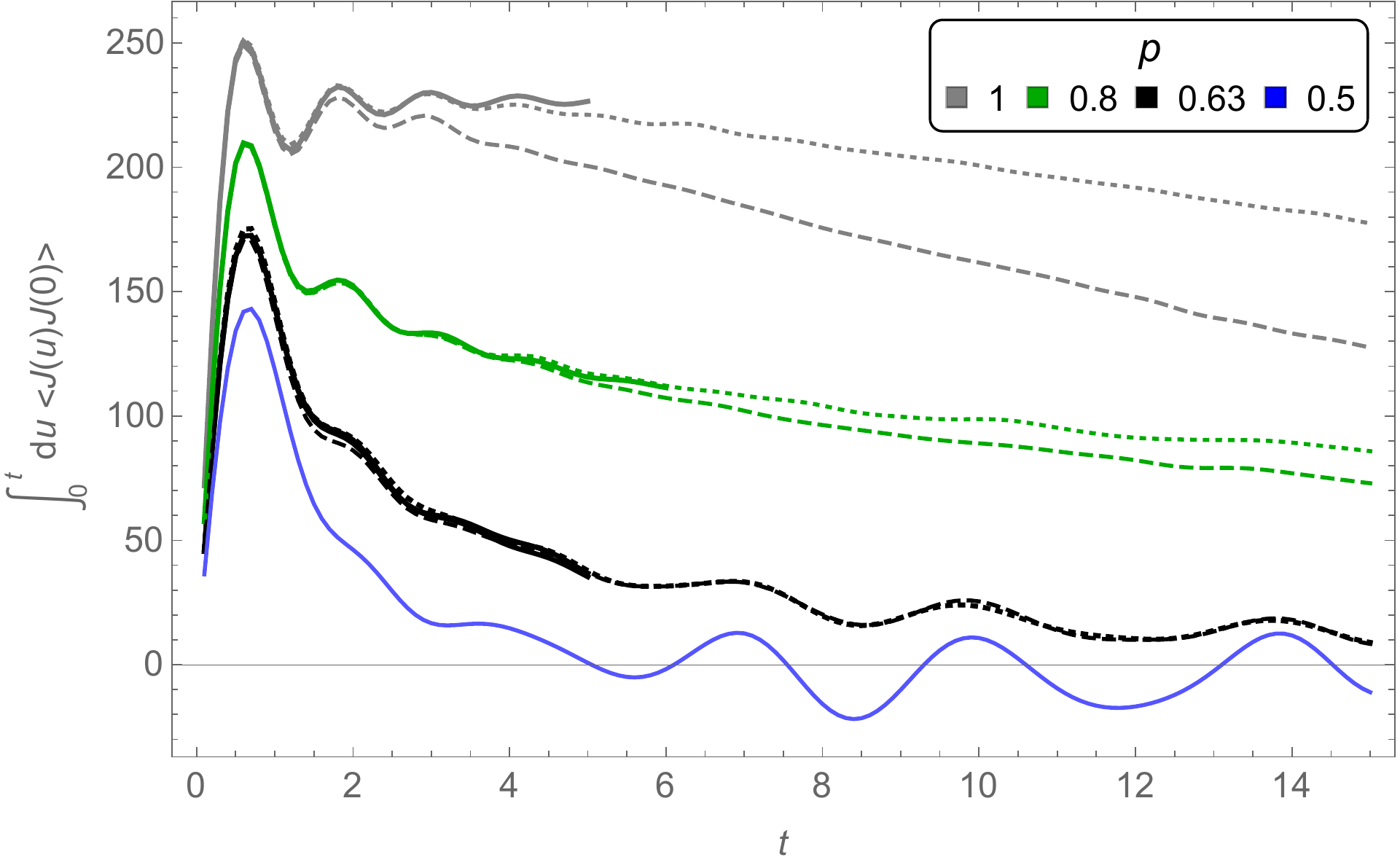}
\caption{Finite-size scaling of apparent conductivity at various system sizes and $p$. Values of $p$ are color-coded as in the legend; within each group, dashed lines represent $N = 50$, dotted lines $N = 100$, and solid lines $N = 400$. All data are averaged over 400 realizations.}
\label{fss}
\end{center}
\end{figure*}

Finally, we discuss the d.c. limit of the conductivity. As one can see from the discussion of the optical conductivity, extrapolating to the d.c. limit from the finite-frequency data is delicate. The right panel of Fig.~\ref{compfig} does this by computing an apparent conductivity that one extracts by integrating the correlation function $\langle J(t') J(0) \rangle$ out to some time $t$. In a diffusive system this integral converges to a finite answer, which is the d.c. conductivity. This is what we see in the unconstrained model. Both the constrained and disordered models, however, have an apparent conductivity that drifts downward as time passes. In the short time window for which we have reliable data on the microscopic model, the two approaches lead to similar conclusions: the constrained value is about $\frac{2}{3}$ of the unconstrained one, and the disordered value is about $\frac{2}{3}$ of the constrained value. However, the conductivity is clearly drifting to lower values as system size increases, suggesting that not all the curvature seen in the diffusion data is due to saturation. We have checked that this behavior is not a finite-size effect in the RRG (Fig.~\ref{fss}). At $p = 1$ the model quickly approaches diffusive behavior, but even for $p = 0.8$ (deep in the percolating phase) the apparent d.c. conductivity drifts down with time and it is unclear whether this quantity is finite. This phenomenon seems related to the existence of critical states, but we do not have a clear understanding of it. 

%
%

\bibliography{BiblioPaper02}

\end{document}